\documentclass[12pt]{article}

\usepackage{amsmath,amssymb,exscale}
\usepackage{array,multicol}
\usepackage{afterpage,float,flafter}
\usepackage{epsfig,rotating,pifont}
\usepackage{cite}
\def\beq{\begin{eqnarray}}
\def\eeq{\end{eqnarray}}

\def\lsim{\mathrel{\rlap{\lower3pt\hbox{\hskip0pt$\sim$}}
    \raise1pt\hbox{$<$}}}         
\def\gsim{\mathrel{\rlap{\lower4pt\hbox{\hskip1pt$\sim$}}
    \raise1pt\hbox{$>$}}}         

\newcommand{\be}{\begin{equation}}
\newcommand{\ee}{\end{equation}}
\newcommand{\bea}{\begin{eqnarray}}
\newcommand{\eea}{\end{eqnarray}}
\newcommand{\bg}{\begin{gather}}
\newcommand{\eg}{\end{gather}}
\newcommand{\bseq}{\begin{subequations}}
\newcommand{\eseq}{\end{subequations}}

\def\tr{\hbox{Tr}}

\def\be{\begin{eqnarray}}
\def\ee{\end{eqnarray}}
\def\lb{\label}

\title{
{\LARGE\textbf{Black Hole Entropy and Gravity Cutoff } }
\vspace{1cm}
\author{
\textbf{Gia Dvali{\boldmath $^{\,a,b}$} and
 Sergey N.
  Solodukhin{\boldmath $^{\,a,c}$}}\\
  \\
$^a$\normalsize\emph{Theory Group, Physics Department, CERN,
CH-1211 Geneva 23,
  Switzerland}\\
$^b$\normalsize\emph{Center for Cosmology and Particle Physics,
Department of Physics,}\\
\normalsize\emph{
New York University, New York NY 10003, USA}\\
$^c$\normalsize\emph{Laboratoire de Math\'ematiques et Physique
Th\'eorique CNRS-UMR 6083, }\\
\normalsize\emph{Universit\'e de Tours, Parc de Grandmont, 37200
Tours, France} }}

\date{}
\begin{document}
\maketitle \thispagestyle{empty}

\vspace{-12cm}
\begin{flushright}
CERN-PH-TH/2008-141
\end{flushright}

\vspace{10cm}

\begin{abstract}
We study  the black hole entropy as entanglement entropy and
propose a resolution to the  species puzzle. This resolution comes
out naturally due to the fact that in the presence of $N$ species the universal gravitational cutoff
is $\Lambda=M_{\rm Planck}/\sqrt{N}$, as opposed to  $M_{\rm Planck}$.
We demonstrate consistency of our solution by showing the equality of the two
entropies  in  explicit examples in which the relation between $M_{\rm Planck}$
and $\Lambda$ is known from the fundamental theory.

\end{abstract}

\newpage
\renewcommand{\thepage}{\arabic{page}}
\setcounter{page}{1}

\section{Introduction}

One of the most important unresolved theoretical problems is the
existence of black holes which quantum mechanically possess
entropy even though  there are no obvious degrees of freedom in
the theory that would constitute this entropy statistical
mechanically. Despite the numerous attempts, including successful
ones proposed in string theory, it remains a mystery what produces
the huge Bekenstein-Hawking (BH) entropy \cite{Bekenstein:1973ur},
\cite{Hawking:1974sw} \be S_{BH}=M^2_{\rm Planck}A(\Sigma)~~,
\lb{0} \ee where $M^2_{\rm Planck}=1/(4G_N)$, of an uncharged
macroscopic 4-dimensional black hole. Among others, the most
intriguing approach to attack this problem is to use entanglement
entropy. Obvious advantage of this  approach is its universality,
the mechanism of generation of the entropy is the same for all
possible types of black holes, and its geometrical nature, the
fact that the BH entropy is geometric finds a simple explanation.
The identification of the BH entropy with entanglement entropy,
however, has a well-known difficulty: the black hole entropy is
universal finite quantity while the entanglement entropy is
quadratically-sensitive to the cutoff of the theory ($\Lambda$)
and grows with the number of particle species $N$. In this note we
suggest that the species problem
results from the use of the wrong cutoff. Namely, the mistake is in
ignoring the $N$-dependence of the cutoff of the theory.
The latter dependence follows from completely independent consistency
arguments, which  show that there is a strict
relation between $\Lambda$, $N$ and $M_{\rm Planck}$, so that with growing
number of species, $\Lambda$ decreases relative to the Planck mass by
$\sqrt{N}$. We show, that when this is taken into the account, the puzzle
of species disappears and the two entropies match to leading order.

\section{Entanglement Entropy of Black Holes}
Entanglement entropy \cite{ent} is defined for a system divided by
surface $\Sigma$ into two subsystems $A$ and $B$. Typically, the
total system is considered to be in a pure state (for instance, in
a ground state) described by wave function $|\Psi>$ so that one
defines $\rho(A,B)=|\Psi ><\Psi |$. If one does not have access to
the modes in one of the subsystems, say $B$, then one should trace
over these modes and this situation is described by the reduced
density matrix $\rho_A=\tr_B \rho(A,B)$. Thus, provided we have
access only to a part of the system then entanglement entropy
defined as $S_A=-\tr_A \rho_A\log \rho_A$ gives us a measure for
the lack of information about the state of the total system. We
could have traced over modes in subsystem A and get density matrix
$\rho_B$ and respective entropy $S_B$. If the total system was in
a pure state then $S_A=S_B$. This property of entanglement entropy
means that if the entropy is non-zero then it should not depend on
extensive quantities (such as  volume) which characterize A or B
but only on the surface $\Sigma$ that separates the two
subsystems.

On the other hand, this entropy is non-vanishing because there are
correlations between  the subsystems.
 Entanglement entropy,
thus, can be also viewed as some measure for these correlations.
In local theories the correlations are short-distance and, hence,
the entropy
 is expected to be determined by the geometry of the boundary dividing the two
subsystems. It is clear that in order to regularize these
correlations one has to introduce a short-distance cut-off
$\epsilon=1/\Lambda$ so that the entropy essentially depends on
the cut-off.

The natural (if not the only one possible) way to prevent the
access to a part of the system inside a closed surface $\Sigma$ is
to put the system into a black hole space-time so that $\Sigma$
would be a black hole horizon. In this situation the access to the
inside region of $\Sigma$ is impossible in principle, so that the
notion of entanglement entropy is well suited to describe the lack
of information in black hole. In order to calculate the entropy
then we have to start with a pure state described by a wave
function of black hole (for a construction of this function see
\cite{Barvinsky:1994jca})  $\Psi(\phi_{in},\phi_{out})$ that is a
functional of the modes inside horizon $\Sigma$ and modes
$\phi_{out}$ outside $\Sigma$. No observer has an access to the complete
set of modes. By tracing over modes $\phi_{in}$ we end up with a
density matrix and the corresponding entropy $S_{\rm ent}$ depends
on the geometry of $\Sigma$. Again the short-distance correlations
are present in the system that should be cured by an UV cutoff
$\epsilon=1/\Lambda$. As far as the technicalities are concerned
\cite{Solodukhin:1994yz}, the calculation  of entanglement entropy
of black hole horizon goes along the same steps as  in the flat
space-time. The only essential difference is that the surface in the
flat spacetime may have non-trivial extrinsic geometry while  the
extrinsic curvature  vanishes for black hole horizons. This
restricts the geometric structures which may appear in the
entropy.
 To leading order entanglement entropy is proportional
to the area $A(\Sigma)$ of $\Sigma$. If there are $N$ fields
available, then each of them equally contributes to the entropy so that
the entanglement entropy of a black hole in four dimensions is given
by
 \be S_{\rm ent}={N\Lambda^2}A(\Sigma)~~.
\lb{1} \ee The fact that the entanglement entropy is proportional to
the area similarly to the Bekenstein-Hawking (BH) entropy makes it an
interesting candidate for the statistical origin of the BH
entropy. A would be natural identification $\Lambda\sim M_{\rm
Planck}$ however faces  a well-known problem of species (see
\cite{Bekenstein:1994bc} for a review): entanglement entropy
depends on the number $N$ of the species while the BH entropy is
obviously universal and should not depend on $N$.
 We now wish to show, that in order to resolve this puzzle
we have to invoke the correct gravity cutoff in the presence of $N$ species\cite{bound}.

\section{Gravity Cutoff}
It was proposed in \cite{bound} that in a theory with $N$ species
the self-consistency of large-distance black hole physics implies the existence of
the following new scale
\be \label{cutoff} \Lambda  \, = \, M_{\rm Planck}/\sqrt{N}\lb{2}\,.
\ee

 The physical meaning of this scale is twofold. First \cite{bound}, it sets
the bound on the masses of the particle species.  The same scale sets the
bound on the gravitational cutoff of the theory\cite{dr}.
We shall now briefly reproduce some of the key consistency arguments from the black hole physics.

 Following \cite{bound}, let us consider a theory with $N$ particle species, $\Phi_j~~
j=1,2,...N$.  For simplicity, we shall assume the species to be stable, and to be having
equal masses, which we denote by $M$.  It is useful to monitor the ``personalities''
of the individual species by prescribing them parities under some gauged $Z_2$-symmetries, one per each species. Under a given $Z_2^{(j)}$-symmetry, only a corresponding species changes the sign,  $\Phi_j \, \rightarrow \, -\, \Phi_j$, whereas the remaining $N-1$ species stay invariant. We then perform the following thought
experiment.  Imagine an arbitrarily large classical black hole, and let us endow
it with a maximal possible discrete charge. This can be done by throwing in
the black hole one particle per each species.
In this way, the resulting black hole will carry $N$ units of different
$Z_2^{(j)}$-charges. Since the symmetries are gauged, this charge cannot be lost and must be returned back after the black hole evaporation. (The discrete gauge charges can be
continuously monitored at infinity, by Aharonov-Bohm type experiments\cite{quantum}).
However, irrespective of the original size, the black hole can only start giving back the charge after its Hawking temperature becomes comparable to the particle masses,
$T_H \, \sim \, M$.  (Emission of particles before this moment is Boltzmann-suppressed
and can only correct our results by factor $\sim$log$N$).
At this point, the black hole mass is $M_{BH} \sim M_{\rm Planck}^2/M$. After this moment, we can use the conservation of energy and immediately derive that the maximal number of quanta that can be emitted by the black hole is  $N_{max} \, \lsim \,
M_{\rm Planck}^2 /M^2$. This proves that the bound on particle masses is set by $\Lambda$, defined by (\ref{cutoff}).

 Having this bound on the particle masses already indicates that the cutoff of the theory also should be somewhere around the same scale $\Lambda$. For example, assume that the species are self-interacting scalars. If cutoff were much higher than the scale
$\Lambda$, it would be hard to reconcile the lightness
of the species versus the cutoff, in the view of quantum corrections to their masses that are quadratically sensitive to the cutoff. And indeed, the cutoff is at the scale $\Lambda$.

 In order to see this, we can again use as a tool the well-known properties of the
black holes \cite{dr}. Let us assume the opposite, that the gravity cutoff is much higher
than the scale $\Lambda$. We shall now see that with this assumption, we shall inevitably run into an inconsistency with the well-established macro black hole physics. If the true gravitational cutoff is much
higher than the scale $\Lambda$, then the black holes of size $\Lambda^{-1}$ must behave as normal Schwarszchild black holes. Thus, consider
such a black hole.  In the normal case, with only few species, the lifetime of a black hole of size $\Lambda^{-1}$ would be
\begin{equation}
 \tau_{BH} \, \sim \, \frac{M_{\rm Planck}^2}{\Lambda^3}\, .
\end{equation}
For $\Lambda \, \ll \, M_{\rm Planck}$, this lifetime would be perfectly consistent with our assumption that such a black hole is a quasi-classical object with well-defined
(slowly-changing) Hawking temperature, $T_{H} \, \sim \, \Lambda$. However, because of
$N$ species, the evaporation rate is enhanced by factor of $N$.  Taking into the account (\ref{2}), this enhancement  reduces the black hole lifetime to
\begin{equation}
\tau_{BH} \, \sim \, \Lambda^{-1}\, .
\end{equation}
Thus, the black hole has a lifetime of order of its inverse temperature! This is a clear indication that such a black hole cannot be regarded as a quasi-classical state, with a well-defined  temperature.  Thus, we are inevitably lead into  the contradiction
with our initial assumption that black holes of size $\Lambda^{-1}$ are
 normal Schwarzschild black holes. The only resolution of this inconsistency is that the
gravity cutoff is $\Lambda$.

 Finally, let us note that the above presented non-perturbative arguments exactly match
 the perturbative ones \cite{dg, Veneziano:2001ah, dr}, which also indicate that
in theory with $N$ species gravitational cutoff must be set by
$\Lambda$, and this is the scale above which the low energy
perturbation theory breaks down\footnote{In \cite{dg} this fact
was used to explain largeness of the four-dimensional Planck scale
relative to the fundamental high-dimensional one, in theory where
$N$ four-dimensional particle species are localized on a 3-brane
embedded in extra space.}.

 Thus, there are number of different perturbative and non-perturbative arguments
for the validity of the relation (\ref{2}), all of which will not be reproduced here.
For more detailed discussions the reader is referred to the above mentioned papers.
The most relevant point for our present subject is the fact that the scale
$\Lambda$ is the gravitational cutoff.

Given the relation (\ref{cutoff}),  our strategy for the  resolution of the species puzzle is clear.
Since the scale (\ref{2}) is the right UV cutoff,
we should use $\Lambda$ and {\it not} $M_{\rm Planck}$
 in the entanglement entropy calculation. Then the
disturbing dependence of the entanglement entropy on the number of
species disappears and the entropy (\ref{1}) precisely agrees with
the BH entropy (\ref{0}).

\section{Explicit Examples}
\subsection{Entropy of High-Dimensional Black Holes}
 We now wish to demonstrate the consistency of our solution on explicit examples,
 in which the relation (\ref{cutoff}) is fixed by the fundamental theory.
 We then demonstrate that in such cases, the two entropies  are automatically equal, despite the
 presence of  the arbitrarily large number of species.

   As such example, consider $4+n$ dimensional theory with  $n$ space dimensions compactified on $n$-torus, and $4$ non-compact dimensions forming Minkowskian geometry.  Without affecting any of our results,  for simplicity, we shall
 set all the radii being equal to $R$.  We shall choose the radius $R$ to be much larger  than the
   fundamental Planck length $\Lambda^{-1}$, and otherwise keep it as a free parameter, which can
  be consistently taken to infinity.   In this limit one recovers a $4+n$ dimensional Minkowski space.
    The fundamental high-dimensional theory  has one
 dimensionfull parameter, the $4+n$-dimensional Planck mass $\Lambda$,  which is the cutoff of the theory.  We shall assume that the only species in the theory is a $4+n$-dimensional graviton.
From the four-dimensional point of view it is a theory of the tower of spin-2 Kaluza-Klein  species.
For any $R$,  the relation between the four-dimensional Planck mass and the cutoff of the theory is\cite{add}
\begin{equation}
\label{planckscales} M_{\rm Planck}^2 \, = \, \Lambda^2 \,
(R\Lambda)^n
\end{equation}
Notice that the factor $(R\Lambda)^n$ measures the number of KK species
\begin{equation}
\label{kknumber}
N \, = \, (R\Lambda)^n \,.
\end{equation}
 Thus,  as already noted in \cite{bound,dr}, the relation (\ref{planckscales}) is a particular example
of relation (\ref{cutoff}) in which $N$ has to be understood as the number of KK species.

 Let us now prove that a well known  Bekenstein-Hawking  entropy of a high-dimensional black hole,
 is correctly reproduced by the entanglement entropy of  $N$ KK species.  For this consider
 a high dimensional black hole of gravitational radius $r_g$.

 First we consider a small black hole for which $R \, \gg \ r_g \gg \Lambda^{-1}$. In  this regime, on one hand
 black hole is classical, and on the other hand the effects of compactification can be ignored
 on the near-horizon geometry. The black hole horizon is
 $(2+n)$-sphere of radius $r_g$ in this case.
 We can then use the well known generalization of Beckenstein-Hawking entropy
 for a  high dimensional  black hole\cite{highd}.  Not surprisingly, this entropy is given by the  black hole area in
 fundamental Planck units
\begin{equation}
\label{entropyn}
 S_{BH}= \, (\Lambda \, r_g)^{2+n}.
\end{equation}
Note that this is entropy of the black hole in the
$(n+2)$-dimensional theory.
 Let us now show that, from 4-dimensional perspective,  this equation can be understood as entanglement entropy of KK species.
We can do this in two ways, working with species that are either  coordinate  or momentum
eigenstates in extra dimensions.  Conventional KK expansion in the flat space
is done in eigenstates of  high-dimensional momentum operator, which coincides  with eigenstates
of four-dimensional mass operator.   In accordance with the usual uncertainty relation, the
wave-functions of KK states are plane waves $\Psi_{\vec{P}} \, =\, $ e$^{i\vec{P}_{extra}\vec{Y}_{extra}}$ in the extra coordinate, forming a complete set of functions.  Cutting of this infinite tower by  $P_{extra} \, \leq \, \Lambda$, we get $N$ KK,  species as explained above.

  By forming appropriate orthogonal superpositions of the momentum eigenstates, we could form a complete set of coordinate eigenstates, each being localized
at one particular point in extra dimensions, $\psi_{\vec{Y}}\, = \, \delta ^n (\vec{Y} - \vec{Y}_0)$
However, since we are limited by the cutoff in momentum space, the delta functions in position space will be smeared over a distance $\sim \Lambda^{-1}$,  and again we get $N$ coordinate eigenstates, localized  at $\Lambda^{-1}$ distance apart.   For each of these localized states we can compute the entanglement entropy. For each state what will matter is the radial (in extra coordinate) distance
 from its localization site to the center of the black hole. In the other words, for each state localized on a
 $3$-dimensional surface in $3+n$ dimensional space, the black hole horizon will cut out a $2$-dimensional sphere of radius
 $r$, where $0 \leq r \leq r_g$.  Suppressing the factors of order one, the
entanglement entropy for each such state will be
 \begin{equation}
\label{individual}
S_{ind} \, = \, (r\Lambda)^2.
\end{equation}
Integrating this over all possible localization sites,  we obviously  get
\begin{equation}
\label{total} S_{ent} \, = \ (r_g\Lambda)^2 \, N(r_g) =\,  (r_g
\Lambda)^{n+2}  \,,
\end{equation}
where $N(r_g)=(r_gN)^n$ is the number of distinct species
localised in the extra n dimensions that see the black hole
horizon of radius $r_g$. This equation  correctly reproduces
(\ref{entropyn}). Despite the existence of arbitrarily large
number of species, no puzzle appears, since $N$ dependence is
automatically taken care of by the relation (\ref{planckscales}).
Had we incorrectly used $M_{\rm Planck}$ as a cutoff we would end
up with the species puzzle.

   We can do the same computation using momentum eigenstates. Each KK species sees the
   in four-dimensions the cut-out
   sphere of surface $r_g^2$,  which is simply a four-dimensional projection of the high-dimensional black hole horizon.    However, because the wave-function of each KK is spread out in extra dimension, we have to take into the account the intersection probability.  Since the species  are the uniform plane waves
in $n$ extra dimensions, whereas the cross-section of the black
hole is $n$ dimensional sphere of volume $r_g^n$, the intersection
probability is equal to $\frac{r_g^n}{  R^n}$. Thus the individual
contributions to the entanglement entropies are equal to
 \begin{equation}
\label{kk} S_{KK} \, = \, (r_g \Lambda)^2 \left ( \frac{r_g}{  R}
\right )^n.
\end{equation}
Summing this over all the KK states,  with total number  $N$ given by (\ref{kknumber}), we get  exactly
(\ref{entropyn})
\begin{equation}
\label{ }
S_{ent} \, = \, \sum_{KK} \, S_{KK}  \, =  \, N \, S_{KK} \, = S_{BH}.
\end{equation}

Consider now the case of large black hole $r_g\gg R$. In this case
  the black hole horizon is a product of 2-sphere
of radius $r_g$ and $n$-dimensional torus of size $R$. From the
point of view of 4-dimensional gravity this black hole has entropy
\be S^{(4)}_{BH}=(M_{\rm Planck} r_g)^2 \lb{entropy4} \ee which,
by means of relation (\ref{planckscales}), is identical to the
Bekenstein-Hawking entropy in  $(4+n)$-dimensional theory \be
S^{(4+n)}_{BH}=\Lambda^{n+2} r_g^2R^n \lb{entropyL} \, ,\ee where
$4\pi r_g^2R^n$ is the area of $(4+n)$-dimensional horizon.

The calculation of the entanglement entropy remains the same.  The
entropy of a single KK state (in the coordinate representation)
has  entropy \be S_{ind}=(r_g\Lambda)^2\, . \lb{sind} \ee The
number of states in the case when $r_g\gg R$ does not depend on
$r_g$ and is equal to $N=(R\Lambda)^n$ (\ref{kknumber}), the full
number of KK species. Summing over all species we get \be S_{\rm
ent}=(r_g\Lambda)^2\, N=(r_g\Lambda)^2(R\Lambda)^n \lb{ente} \ee
which exactly agrees both with (\ref{entropy4}) and
(\ref{entropyL}).

That the entanglement entropy calculation gives correct entropy in
two limiting cases ($R\gg r_g$ and $r_g\gg R$) makes us beleive
that this method should work also in the intermediate regime
although to do the computation one should know the concrete
profile of the black hole horizon along the extra dimensions.

\subsection{AdS/CFT Example} To further illustrate our point and
provide the reader with another explicit example we consider a
3-brane in a $Z_2$ configuration in anti-de Sitter space-time.
This is the Randall-Sundrum set-up, in the framework of the
AdS/CFT correspondence it has a description in terms of CFT on the
brane coupled to gravity at a UV cutoff \cite{3brane}. If the
brane is placed at the distance $\rho=\epsilon^2$ from
the Anti-de Sitter boundary, one obtains that there is a dynamical gravity
induced on the brane with the induced Newton constant
$1/G_N={2N^2/(\pi \epsilon^2)}$. According to the AdS/CFT
dictionary,  in the theory on the brane
$\Lambda={1/\sqrt{2\pi}\epsilon}$ has the meaning of the UV
cut-off. Thus, one finds that $\Lambda=M_{\rm Planck}/N$, in
agreement with (\ref{2}). We remind that the quantum field theory
defined on the brane is the super-conformal $SU(N)$ gauge theory,
so that in the large $N$ limit $N^2$ represents the number of
species. Thus, the 3-brane in the anti-de Sitter space-time gives
us an example when the relation (\ref{2}) holds automatically.
Consider now a black hole on the 3-brane and compute its
entanglement entropy \cite{ads/cft} due to the CFT. To leading
order one finds
 \be
S_{\rm ent}=\frac{N^2}{4\pi
\epsilon^2}A(\Sigma)={N^2\Lambda^2}A(\Sigma) , \lb{3} \ee which is
exactly the Bekenstein-Hawking entropy of the black hole provided
one expresses the UV cut-off in terms of the induced Planck scale
$M_{\rm Planck}=N\Lambda$ .

The fact that the BH entropy is correctly reproduced in the
entanglement entropy calculation, in accordance to our proposal, is
not limited to the above particular examples, but is much more general
 and relies solely on the existence of the cutoff (\ref{2}).

\section{Conclusions}

In this note we consider the black hole entropy as entanglement
entropy and  propose a resolution to the species puzzle. We
suggest that the puzzle never appears provided the correct cutoff
in the theory is $\sqrt{N}$ suppressed relative to the Planck
mass. This suppression follows from completely independent
consistency arguments given in \cite{bound}. We demonstrate the
equality of two entropies in explicit examples in which the
relation between the Planck mass and the cutoff is known from the
fundamental theory. It would be interesting to verify our proposal
directly in string theory. For this one would have to compute
entanglement entropy of strings (for a recent work in this
direction see \cite{Brustein:2005vx}).

It is straightforward to generalize our proposal to arbitrary
dimensions. Indeed, the BH entropy of a black hole in $d$
dimensions, $S_{\rm BH}^{(d)}=M^{d-2}_{(d)}A(\Sigma)$, where
$M_{(d)}$ is the Planck mass in d-dimensional gravity and
$A(\Sigma)$ is the area of $(d-2)$-surface of the horizon, matches
the entanglement entropy $S^{(d)}_{\rm
ent}=N\Lambda^{d-2}A(\Sigma)$ produced by $N$ species at cutoff
$\Lambda$ if the relation between $M_{(d)}$, $N$ and $\Lambda$ is
given by \be \Lambda^{d-2}N=M^{d-2}_{(d)}\, .\lb{ddim}\ee This is
exactly the generalized gravity cutoff in $d$ dimensions found in
\cite{dr}. Thus, our proposal resolves the species problem  in
arbitrary dimensions.

\subsection*{Acknowledgments}

The work  is supported in part
by David and Lucile  Packard Foundation Fellowship for  Science
and Engineering, and by NSF grant PHY-0245068. S.S. is grateful to
the Theory Division at CERN for the hospitality extended to him
while this work was in progress.



\begin{thebibliography}{99}


\bibitem{Bekenstein:1973ur}
  J.~D.~Bekenstein,
  Phys.\ Rev.\  D {\bf 7} (1973) 2333.

\bibitem{Hawking:1974sw}
  S.~W.~Hawking,
  Commun.\ Math.\ Phys.\  {\bf 43}, 199 (1975).

\bibitem{ent} W. Israel, Phys. \ Lett.\ {A57} (1976) 107;\\
  L.~Bombelli, R.~K.~Koul, J.~H.~Lee and R.~D.~Sorkin,
  %
  Phys.\ Rev.\ D {\bf 34}, 373 (1986);\\
  M.~Srednicki,
  %
  Phys.\ Rev.\ Lett.\  {\bf 71}, 666 (1993);\\
  V.~P.~Frolov and I.~Novikov,
  Phys.\ Rev.\ D {\bf 48}, 4545 (1993).

\bibitem{bound}
 G.~Dvali, {\em
 Black Holes and Large N Species Solution to the Hierarchy Problem},\\{}
  arXiv:0706.2050 [hep-th].



\bibitem{dr}
G.~Dvali and M.~Redi,
  Phys.\ Rev.\  D {\bf 77}, 045027 (2008)
  [arXiv:0710.4344 [hep-th]].

 \bibitem{quantum}
  L.M.~Krauss and F.~Wilczek,  {\it Phys. Rev. Lett.} {\bf 62} (1989)
  1221.\\
  J.~Preskill and L.M.~Krauss, {\it Nucl. Phys.} {\bf B341} (1990)
  50.\\
 S.~Coleman, J.~Preskill and F.~Wilczek, {\it Phys. Rev. Lett.} {\bf 67} (1991) 1975.

\bibitem{dg}
G.~Dvali and G.~Gabadadze,
Phys.\ Rev.\ D {\bf 63}, 065007 (2001) [hep-th/0008054].

\bibitem{Veneziano:2001ah}
  G.~Veneziano,
  JHEP {\bf 0206}, 051 (2002)
  [arXiv:hep-th/0110129].

\bibitem{add}
N.~Arkani-Hamed,  S.~Dimopoulos and G.~Dvali, {\it Phys. Lett.}
{\bf B 429} (1998) 263, hep-ph/9803315. {\it Phys. Rev.} {\bf D
59} (1999) 086004, hep-ph/9807344.\\
 I.~Anatoniadis,
N.~Arkani-Hamed, S.~Dimopoulos and  G.~Dvali, {\it Phys. Lett.}
{\bf 436} (1998), hep-th/9804398.

\bibitem{Barvinsky:1994jca}
  A.~O.~Barvinsky, V.~P.~Frolov and A.~I.~Zelnikov,
  Phys.\ Rev.\  D {\bf 51}, 1741 (1995).

\bibitem{Solodukhin:1994yz}
  S.~N.~Solodukhin,
  %
  Phys.\ Rev.\ D {\bf 51}, 609 (1995);
 [arXiv:hep-th/9407001].\\
S.~N.~Solodukhin,
  ``Entanglement entropy, conformal invariance and extrinsic geometry,''
  arXiv:0802.3117 [hep-th].

\bibitem{Bekenstein:1994bc}
  J.~D.~Bekenstein,
  {\em Do we understand black hole entropy?,}
  arXiv:gr-qc/9409015.

\bibitem{highd}
  R.~C.~Myers and M.~J.~Perry,
  Annals Phys.\  {\bf 172}, 304 (1986).

\bibitem{3brane}
  L.~Randall and R.~Sundrum,
  Phys.\ Rev.\ Lett.\  {\bf 83}, 4690 (1999)
  [arXiv:hep-th/9906064]; \\
  S.~S.~Gubser,
  Phys.\ Rev.\  D {\bf 63}, 084017 (2001)
  [arXiv:hep-th/9912001].


\bibitem{ads/cft}
  S.~Hawking, J.~M.~Maldacena and A.~Strominger,
  JHEP {\bf 0105}, 001 (2001);
  [arXiv:hep-th/0002145].\\
S.~N.~Solodukhin,
  Phys.\ Rev.\ Lett.\  {\bf 97}, 201601 (2006);
 [arXiv:hep-th/0606205].\\
  R.~Emparan,
  JHEP {\bf 0606}, 012 (2006);
 [arXiv:hep-th/0603081].

\bibitem{Brustein:2005vx}
  R.~Brustein, M.~B.~Einhorn and A.~Yarom,
  JHEP {\bf 0601}, 098 (2006)
  [arXiv:hep-th/0508217].
\end{thebibliography}
\end{document}